\documentclass[aps,prd,showpacs,onecolumn,preprintnumbers,notitlepage,
nofootinbib,amsfont,amssymb,10pt,singlespacing]{revtex4-1}
\usepackage{amsmath}
\usepackage{bm}
\usepackage{times}
\usepackage{braket}
\usepackage{color}
\usepackage{epsfig}
\usepackage{slashed}
\usepackage{hyperref}

\newcommand{\beq}{\begin{eqnarray}}
\newcommand{\eeq}{\end{eqnarray}}
\newcommand{\non}{\nonumber\\ }

\def\lsim{ {\ \lower-1.2pt\vbox{\hbox{\rlap{$<$}\lower6pt\vbox{\hbox{$\sim$}
}}}\ } }
\def\gsim{ {\ \lower-1.2pt\vbox{\hbox{\rlap{$>$}\lower6pt\vbox{\hbox{$\sim$}
}}}\ } }

\def \jhep{ J. High Energy Phys.  }

\definecolor{Red}{rgb}{1.,0.,0.}

\definecolor{Blue}{rgb}{0.,0.,1.}

\definecolor{nicered}{rgb}{0.7,0.1,0.1}
\definecolor{nicegreen}{rgb}{0.1,0.5,0.1}

\hypersetup{colorlinks,citecolor=nicegreen,linkcolor=nicered}

\begin{document}

\title{\boldmath Nonleptonic charmless decays of $B_c\to TP, TV$
in the perturbative QCD approach}
\author{Xin~Liu}
\email[Electronic address: ]{liuxin@jsnu.edu.cn}
\affiliation{School of Physics and Electronic Engineering, Jiangsu Normal University, Xuzhou 221116, China}

\author{Run-Hui~Li}
\email[Electronic address: ]{lirh@imu.edu.cn}
\affiliation{School of Physical Science and Technology,
Inner Mongolia University, Hohhot 010021, China}

\author{Zhi-Tian~Zou}
\email[Electronic address: ]{zouzt@ytu.edu.cn}
\affiliation{Department of Physics, Yantai University, Yantai 264005, China}

\author{Zhen-Jun~Xiao}
\email[Electronic address: ]{xiaozhenjun@njnu.edu.cn}
\affiliation{Department of Physics and Institute of Theoretical
Physics, Nanjing Normal University, Nanjing 210023, China}


\date{\today}

\begin{abstract}

Two-body charmless hadronic $B_c$ decays involving a light $1^3\!P_2$-tensor($T$) meson are investigated for the first time
within the framework of perturbative QCD(pQCD) 
at leading order, in which the other meson is the 
lightest
pseudoscalar($P$) or vector($V$) state. The concerned processes
can only occur through the pure weak annihilation topology in
the standard model. We predict the {\it CP}-averaged branching
ratios and polarization fractions of those considered
decays in Cabibbo-Kobayashi-Maskawa(CKM) favored and suppressed
modes. Phenomenologically, several modes$-$such as the
$B_c \to K_2^*(1430) K$ and the CKM-favored $B_c \to TV-$have 
large decay rates of $10^{-6}$, which are expected to
be detected at 
Large Hadron Collider experiments in the
near future. Moreover, all of the $B_c \to TV$ modes are
governed by the longitudinal amplitudes in the pQCD
calculations and the corresponding fractions vary
around $78\%$-$98\%$. A confirmation of these
results could prove the reliability of the pQCD
approach used here and further shed some light
on the annihilation decay mechanism.

\end{abstract}


\preprint{\footnotesize JSNU-PHY-HEP-2017-1}
\maketitle

Heavy flavor physics has 
played an 
important role in the
precision tests of the standard model(SM), as well as in
investigating the properties of involved light hadrons
after the advent of two $B$ factories, i.e., {\it BABAR}
at SLAC and Belle at KEK.
An increasing number of interesting mesons
have been observed in the decay channels of the heavy 
mesons$-$specifically, $D_{(s)}$ mesons with a $c$ quark and $B_{(s)}$ mesons
with a $b$ quark~\cite{Olive:2016xmw}$-$which provide a fertile ground
for probing the perturbative and nonperturbative QCD dynamics in the SM. 
With the advent 
of the Large Hadron Collider(LHC) at CERN,
a new territory
has been developed since a great number of $B_c$ meson events
can be observed. 
The properties of the $B_c$ meson and the dynamics
involved in 
$B_c$ decays could 
be fully exploited through
the precision measurements at the LHC with its high collision
energy and high luminosity. Therefore, the $B_c$ meson
decays will open a window to richer physics, which could start 
a new golden era of 
heavy flavor physics with the LHC experiments~\cite{Xiao:2013lia,Liu:2009qa}.

Tensor mesons with quantum number $J^P=2^+$ have recently become
a hot topic. On the one hand, experimentally, {\it BABAR}
and Belle have measured several charmless hadronic $B$ decays
involving a light tensor meson in the final
states~\cite{Olive:2016xmw,Amhis:2016xyh}. Furthermore,
the measurements on the polarization fractions of
$B \to \phi K_2^*(1430)$ decays showed that these two
modes are dominated by the longitudinal polarization
amplitudes, which is 
contrary to the same
$b \to s\bar s s$-transition-induced $B \to \phi K^*$
processes. This phenomenology makes the well-known
``polarization puzzle" more confusing. On the other hand,
theoretically, the tensor meson cannot be produced 
through either local vector or axial-vector operators, or via
the tensor current, which implies that large nonfactorizable
amplitudes or annihilation diagrams would contribute to
the tensor meson emitted modes with experimentally sizable
branching ratios and the relevant investigations should go
beyond the naive factorization. Of course, the polarization
studies on the tensor-vector, tensor-axial-vector, and even
tensor-tensor modes in 
heavy flavor decays can further
shed light on the underlying helicity structure of the decay
mechanism~\cite{Cheng:2010yd}. According to the counting rule,
the annihilation contributions are usually power suppressed,
compared to other spectator diagrams. Nevertheless,
the annihilation contributions are not negligible
and the size is still an important issue in $B$ meson physics (see, e.g.,
Refs.~\cite{Keum:2000ph,Lu:2000em,Lu:2002iv,Li:2004ep,
Kagan:2004uw,Li:2004mp,Hong:2005wj,
Ali:2007ff,Chay:2007ep,Liu:2009qa,Xiao:2011tx,Xiao:2013lia,
Zhu:2011mm,Chang:2014rla,Zou:2015iwa}).
Indeed, the experiments have confirmed some large annihilation decay modes, 
for example, the well-known $B_d \to K^+ K^-$ and $B_s \to \pi^+ \pi^-$ 
decays~\cite{Aaij:2016elb}. Moreover, phenomenologically, the theoretical 
studies on the $B \to \phi K^*$ ~\cite{Kagan:2004uw,Li:2004mp,Zou:2015iwa} 
and $B \to \phi K_2^*(1430)$ decays~\cite{Cheng:2010yd,Kim:2013cpa} have provided  
important improvements in the explanation of the ``polarization puzzle" by including the annihilation effects, though the authors claimed that $f_L(B_d \to \phi K_2^*(1430)) \sim {\cal O}(1)$ with or without the annihilation effects~\cite{Chen:2007qj}.

Compared to the annihilation amplitudes in the charmless $B$ decays, the magnitude 
in the $B_c$ decays would be roughly enlarged by a factor $|V_{cb}/V_{ub}| \sim 11.5$, 
which would consequently result in a 100 times enhancement to the branching ratios. 
Therefore, 
the annihilation $B_c$ modes could possibly 
provide a promising and more appropriate platform 
to study the contributions from the annihilation diagrams, and even further uncover the annihilation decay mechanism. It is great to find that the measurements on the pure annihilation $B_c$ decay modes have been initiated by the LHCb Collaboration, for example, $B_c \to K^+ \bar{K}^0$~\cite{Aaij:2013fja}, $B_c \to K^+ K^- \pi^+$~\cite{Aaij:2016xas}, and $B_c \to p \bar p \pi^+$~\cite{Aaij:2016xxs}, etc. Certainly, with the increasing number of $B_c$ events being collected, more and more annihilation types of $B_c$ decay channels will be opened. Sequentially, much more information on the annihilation decay mechanism must be obtained.

To date, an 
agreement on how to calculate the Feynman diagrams with annihilation topology reliably has not been achieved among the theorists. At least, the perturbative QCD(pQCD)
approach~\cite{Keum:2000ph,Lu:2000em,Li:2003yj} and 
soft-collinear effective theory(SCET)~\cite{Bauer:2000yr}, as two popular tools 
for calculating hadronic matrix elements based on QCD dynamics,~\footnote{Another 
popular method is the QCD factorization approach~\cite{Beneke:1999br,Du:2000ff}, 
which cannot make effective calculations on the annihilation diagrams since there 
exist end-point singularities in the integrals. However, 
data fitting has been broadly adopted in this approach to make theoretical predictions
in the $B_{(s)}$ decays; see, for example,
Refs.~~\cite{Beneke:1999br,Du:2000ff,Beneke:2003zv,Cheng:2009cn,Li:2003hea}} have
rather different viewpoints: the almost imaginary annihilation amplitudes with a 
large strong phase obtained through keeping the parton's transverse momentum in 
the pQCD framework~\cite{Chay:2007ep}, and 
the almost real annihilation amplitudes with a tiny strong phase  obtained by considering the zero-bin subtraction in the SCET framework~\cite{Arnesen:2006vb}.
However, objectively speaking, the confirmation of the predicted branching ratios for the pure annihilation $B_d \to K^+ K^-$ and $B_s \to \pi^+ \pi^-$ decays provided by the CDF~\cite{Ruffini:2013jea} and LHCb~\cite{Aaij:2012as,Aaij:2016elb} collaborations 
provided 
firm
support to the current pQCD approach. 

In this work, we will study the two-body nonleptonic charmless
$B_c$ decays involving a light tensor meson($T$) and a light
pseudoscalar($P$) or vector meson($V$) in the final states
by employing the pQCD approach at leading order.
These considered decays
can only occur through weak annihilation interactions in the SM. Here, the light pseudoscalar(vector) meson includes $\pi$, $K$, $\eta$,
and $\eta'$($\rho$, $K^*$, $\omega$, and $\phi$). 
In the quark model, the observed light tensor meson contains
the isovector states $a_2(1320)$, the isodoublet states
$K_2^*(1430)$, and the isoscalar singlet states $f_2(1270)$
and $f_2^{\prime}(1525)$, 
which have been well
established in various processes~\cite{Olive:2016xmw}.
Hereafter, for the sake of simplicity, we will adopt
$a_2$, $K_2^*$, $f_2$, and $f_2^{\prime}$ to denote
the light tensor mesons correspondingly, unless otherwise
specified. It is worth mentioning that, just like
the $\eta-\eta^{\prime}$ mixing in the pseudoscalar
sector, the two isoscalar tensor states $f_2(1270)$ and
$f_2^{\prime}(1525)$ also have a mixing as
 \beq
\left(
\begin{array}{c} f_2(1270)\\ f_2^{\prime}(1525) \\ \end{array} \right ) &=&
  \left( \begin{array}{cc}
 \cos{\phi_{f_2}} & -\sin{\phi_{f_2}} \\
 \sin{\phi_{f_2}} & \cos{\phi_{f_2}} \end{array} \right )
 \left( \begin{array}{c}  f_{2q}\\ f_{2s} \\ \end{array} \right )\;,
 \label{eq:mix-f1q-f1s}
 \eeq
with $f_{2q} \equiv (u\bar u + d\bar d)/\sqrt{2}$ and $f_{2s} \equiv s\bar s$. The angle between $f_2(1270)$ and
$f_2^{\prime}(1525)$ mixing should be small due to a fact that the former(latter) 
predominantly decays into $\pi\pi (K\bar K)$~\cite{Olive:2016xmw}. Specifically, 
the mixing angle $\phi_{f_2}$ lies in the range 
$6^\circ$-$10^\circ$~\cite{Olive:2016xmw,Li:2000zb,Cheng:2011fk}. Therefore, analogous to $\omega$ and $\phi$ mesons in the vector sector, we will first approximately assume $f_2(f_2^{\prime})$ as the pure $f_{2q}(f_{2s})$ state firstly. The mixture of $f_2-f_2^{\prime}$ with the mixing angle $\phi_{f_2}$ will be left for future studies associated with experimentally precise measurements.

As mentioned above, 
the pQCD approach is an appropriate tool
to effectively calculate the hadronic matrix elements of
annihilation topology in the nonleptonic weak $B$ meson
decays. The most important feature of the pQCD approach
is that it picks up the intrinsic transverse momentum $k_T$
of the valence quarks in light of the end-point divergences
that exist 
in the collinear factorization. Then, based on the
$k_T$ factorization theorem, by utilizing the technique
of resummation the double logarithmic divergences
factored out from the  hard part can be grouped into a
Sudakov factor($e^{-S}$)~\cite{Botts:1989kf} and 
a threshold factor [$S_t(x)$]~\cite{Li:2001ay}, 
which consequently make the pQCD
approach more self-consistent. Then, 
the single logarithmic divergences separated from the hard kernel can be 
reabsorbed into the meson wave functions  
using the eikonal approximation~\cite{Li:1994iu}. The interested reader can refer to the review paper~\cite{Li:2003yj} for more details about this approach. Presently, many quantitative annihilation-type-diagram calculations have been made with this pQCD approach.

\begin{figure}[htb]
\begin{center}
\begin{tabular}{c}
\includegraphics[width=0.7\textwidth]{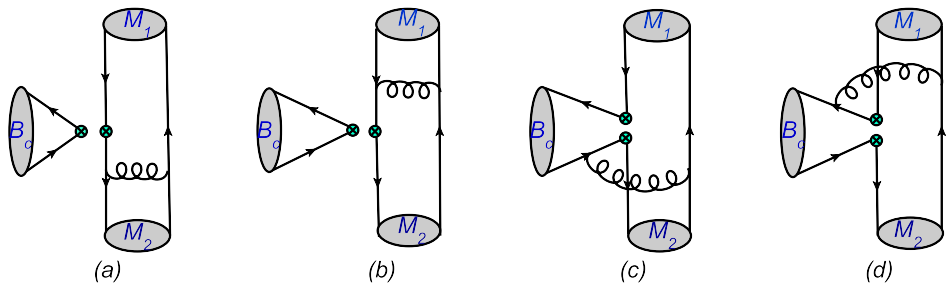}
\end{tabular}
\caption{ Typical Feynman diagrams contributing to
charmless decays of $B_c \to M_1 M_2$ in the pQCD approach
at leading order, where the $M_1M_2$ pair denotes $TP, PT, TV,$ and $VT$ in
this work. }
  \label{fig:fig1}
\end{center}
\end{figure}

The Feynman diagrams for the nonleptonic charmless $B_c \to TP, TV$
decays in the pQCD
approach at leading order 
 are illustrated in Fig.~\ref{fig:fig1}: Figs.~\ref{fig:fig1}(a)
and~\ref{fig:fig1}(b) use 
the factorizable annihilation
topology, while Figs.~\ref{fig:fig1}(c) and~\ref{fig:fig1}(d) use 
 the nonfactorizable annihilation topology. 
For a spin-2 tensor meson, the polarization can be specified by a
symmetric and traceless tensor  $\epsilon^{\mu\nu}_{(\lambda)}$
with helicity $\lambda$ that satisfies the relation $\epsilon_{(\lambda)}^{\mu\nu} P_\mu = \epsilon_{(\lambda)}^{\mu\nu} P_\nu =0$, with $P$ being its momentum. Furthermore, this polarization tensor can be constructed through the spin-1 polarization vector $\epsilon_V$~\cite{Berger:2000wt}. Although a tensor meson contains five spin degrees of freedom, only $\lambda=0$ will give a nonzero contribution in the $B_c \to TP$ modes, since the mother $B_c$ meson is spinless and the daughter $T$ and $P$ mesons should obey the conservation law of angular momentum. Likewise, the $B_c \to TV$ decays will be contributed from $\lambda =0$ and $\lambda =\pm 1$ helicities. Then, one can intuitively postulate that the considered $B_c \to TP, TV$ decays appear more like $B_c \to VP, VV$ ones by elaborating a new polarization vector $\epsilon_T$ for the tensor meson~\cite{Datta:2007yk,Wang:2010ni}.
Actually, $\epsilon_T$ has been explicitly presented in the literature (see, for example, Refs.~\cite{Datta:2007yk,Wang:2010ni,Cheng:2010yd}) with
$\epsilon_T(L) = \sqrt{\frac{2}{3}} \epsilon_V(L)$ and
$\epsilon_T(T)= \sqrt{\frac{1}{2}} \epsilon_V(T)$.~\footnote{Since
only three helicities $\lambda = 0,
\pm 1$ contribute to the $B_c \to TV$ decays, 
the involved light tensor meson can be treated
as a vector-like meson with tensor meson mass.}
Here, the capital $L$ and $T$ in the parentheses describe the 
longitudinal and transverse polarizations, respectively 
( not to be confused
with the abbreviation $T$ for the light tensor meson). The decay
amplitudes of  $B \to TP$ and $B \to TV$ modes presented in
Refs.~\cite{Cheng:2010yd,Kim:2013cpa,Zou:2012td} have confirmed the above postulation.
Therefore, the decay amplitudes
of the $B_c \to TP, TV$ decays considered in this work can be
straightforwardly obtained by replacing the polarization
vector $\epsilon_V$ of the vector meson with the
corresponding $\epsilon_T$ of the tensor one in
the $B_c \to VP, VV$~\cite{Liu:2009qa} modes.
That is:
\begin{itemize}
\item[]{(1)}
Equations~$(28)-(31)$ in Ref.~\cite{Liu:2009qa}
with a factor $\sqrt{\frac{2}{3}}$ will give the analytic
Feynman amplitudes of the $B_c \to PT, TP$ decays with only longitudinal polarization, in which the vector meson mass and distribution amplitudes should be replaced with the tensor state.

\item[]{(2)}
Equations~(49)$-$(50)[Eqs.~(51)$-$(54)] in Ref.~\cite{Liu:2009qa} with a factor $\sqrt{\frac{2}{3}} [\sqrt{\frac{1}{2}}]$ can contribute
to the analytic Feynman amplitudes of the $B_c \to VT, TV$ decays
in longitudinal[transverse] polarizations, where the corresponding quantities of the tensor state will be substituted for those of one of the two vector mesons.
\end{itemize}
Because no $f_2-f_2^\prime$ mixing
is considered, the $B_c \to \pi^+ f_2^\prime$ and $B_c \to
\rho^+ f_2^\prime$  decays will naturally be absent.
Also forbidden is the $B_c \to a_2^+ \phi$ mode as a result
of not including $\omega-\phi$ mixing effects.
Therefore, here we will not
present the factorization formulas and the expressions for
total decay amplitudes explicitly for those considered decays.
The readers can refer to Ref.~\cite{Liu:2009qa} for details.

Now we can turn to the numerical calculations of these
$B_c \to TP, TV$ decays in the pQCD
approach. Before proceeding, some comments on the input quantities are in order:
\begin{itemize}
\item[]{(1)}
For the lightest pseudoscalar and vector mesons, the (chiral) masses, the decay constants, the QCD scale, and the light-cone distribution amplitudes including Gegenbauer moments are same as those used in~\cite{Liu:2009qa}. Please refer to the Appendix A of Ref.~\cite{Liu:2009qa} for detail.

\item[]{(2)}
For the $B_c$ meson, the distribution amplitude and the decay constant are the same as those adopted in Ref.~\cite{Liu:2009qa} but with the up-to-date mass $m_{B_c}=6.275$~GeV and lifetime $\tau_{B_c}=0.507$~ps, which have been updated in the latest version of the Review of Particle Physics~\cite{Olive:2016xmw}.

\item[]{(3)}
For the Cabibbo-Kobayashi-Maskawa(CKM) matrix elements, we also adopt the Wolfenstein parametrization at leading order, but with the updated parameters $A=0.811$ and $\lambda=0.22506$~\cite{Olive:2016xmw}.

\item[]{(4)}
For the light tensor meson, the decay constants with
longitudinal and transverse polarizations are 
collected in Table~\ref{tab:m&f-tensor}.
\begin{table}[hbt]
 \caption{ Decay constants of the light tensor
mesons (in GeV)~\cite{Cheng:2010hn}}
\label{tab:m&f-tensor}
 \begin{center}\vspace{-0.3cm}{
\begin{tabular}[t]{cccccccc}
\hline\hline
$f_{a_2}$ & $f_{a_2}^T$ & $f_{K_2^*}$  &$f_{K_2^*}^T$
&$f_{f_2}$ &$f_{f_2}^T$ & $f_{f_2^\prime}$ & $f_{f_2^\prime}^T$  \\
$0.107\pm 0.006$&$0.105\pm 0.021$
&$0.118\pm 0.005$&$0.077\pm 0.014$
&$0.102\pm 0.006$&$0.117\pm 0.025$
&$0.126\pm 0.004$&$0.065\pm 0.012$\\
\hline \hline
\end{tabular}}
\end{center}
\end{table}
The related masses are $m_{a_2}=1.318$~GeV, $m_{K_2^*}=1.426$~GeV,
$m_{f_2}=1.275$~GeV, and $m_{f_2^\prime}=1.525$~GeV. 
Again, the $f_2-f_2^\prime$ mixing is
not considered in this work. Therefore, the mass of the
pure flavor $f_{2q} (f_{2s})$ state is taken as that of
the physical $f_2 (f_2^\prime)$ 
for convenience.

The related light-cone distribution amplitudes have been
recently investigated in the QCD sum rules~\cite{Cheng:2010hn}.
Analogous to the light vector meson, the asymptotic forms of the
tensor meson distribution amplitudes are adopted.
Here, we present the expressions of the light-cone
distribution amplitudes for the light tensor mesons following Ref.~\cite{Wang:2010ni}: 
\beq
\phi_{T}(x)&=&\frac{3f_{T}}{\sqrt{2N_c}} \phi_\parallel(x)\;,\qquad
\phi_{T}^T(x)=\frac{3f^T_{T}}{\sqrt{2N_c}} \phi_\perp(x)\;,\\
\phi_{T}^t(x)&=&\frac{f_{T}^T}{2\sqrt{2N_c}} h_\parallel^t(x)\;,\qquad
\phi_{T}^s(x)=\frac{f_{T}^T}{4\sqrt{2N_c}} \frac{d}{dx}h_\parallel^s(x)\;,\\
\phi_{T}^v(x)&=&\frac{f_{T}}{2\sqrt{2N_c}} g_\parallel^v(x)\;,\qquad
\phi_{T}^a(x)=\frac{f_{T}}{8\sqrt{2N_c}} \frac{d}{dx}g_\perp^a(x)\;,
\eeq
with
\beq
\phi_\parallel(x)&=&\phi_\perp(x)=x(1-x)[a_1\;C_1^{3/2}(t)]\;,
\label{eq:twist-2} \\
h_\parallel^t(x)&=& \frac{15}{2}(1-6x+6x^2)t, \qquad
h_\parallel^s(x)= 15x(1-x)t\;,\\
g_\perp^v(x)&=&5t^3, \qquad g_\perp^a(x)=20x(1-x)t\;,
\eeq
where the Gegenbauer moment $a_1=\frac{5}{3}$ for the first rough estimates and the Gegenbauer polynomial $C_1^{3/2}(t)=3t$ with
$t=2x-1$. It is worth commenting that, in principle, the Gegenbauer moments 
for different meson distribution amplitudes should usually be different 
due to the expected $SU(3)$ flavor symmetry-breaking effects. Therefore, the larger Gegenbauer moment $a_1$ adopted here will demand further improvements resultant from the near-future relevant measurements with good precision.
\end{itemize}

The pQCD predictions of the {\it CP}-averaged branching ratios in the $B_c \to TP, TV$ decays and of the polarization fractions in the $B_c \to TV$ modes 
collected in Tables~\ref{tab:bc2tp},~\ref{tab:bc2tv0}, and~\ref{tab:bc2tv1}, respectively.
\begin{itemize}
\item[]{(1)}
For the $B_c \to TP$ decays, the main four errors arise from the uncertainties of the 
charm-quark mass $m_c=1.5 \pm 0.15$~GeV in the $B_c$ meson distribution amplitude, of 
the combined decay constants of $f_{T}$ and $f_T^T$ in the tensor meson distribution 
amplitudes, of the combined Gegenbauer moments $a_1$ and/or $a_2$ in the pseudoscalar 
meson distribution amplitudes, and of the chiral mass $m_0^P$ of the pseudoscalar 
mesons.~\footnote{In order to estimate the theoretical uncertainties induced by the 
meson chiral mass, here we consider 10\% variations of the central values for simplicity.} 
Of course, for the $B_c \to TP$ modes involving $\eta$ and $\eta'$ states, we also take the the variations of the mixing angle $\phi_{P}=39.3^\circ \pm 1.0^\circ$ into account as the fifth error.

\item[]{(2)}
For the $B_c \to TV$ channels, the major four errors are induced by the uncertainties of the charm-quark mass $m_c=1.5 \pm 0.15$~GeV in the $B_c$ meson distribution amplitude, of the combined decay constants of $f_T$ and $f_T^T$ in the tensor meson distribution amplitudes, of the combined decay constants $f_V$ and $f_V^T$ in the vector meson distribution amplitudes, and of the combined Gegenbauer moments of $a_1^{\parallel(\perp)}$ and/or $a_2^{\parallel(\perp)}$ in the vector meson distribution amplitudes.
\end{itemize}

Here, we will specify the decay modes into two types: 
the CKM-favored channels with $\Delta S=0$(no strange or two strange
mesons in the final states) and the CKM-suppressed modes with
$\Delta S =1$(only one strange meson in the final states) for
clarifications.
\begin{table}[hbt]
\caption{ {\it CP}-averaged branching ratios of charmless decays $B_c \to TP$ in the pQCD approach. }
\label{tab:bc2tp}
 \begin{center}\vspace{-0.3cm}{
\begin{tabular}[t]{c|c||c|c}
\hline  \hline
   Decay Modes($\Delta S=0$)   &  Branching ratios($10^{-7}$) &  Decay modes($\Delta S=1$) & Branching ratios($10^{-8}$) \\
   \hline \hline
  $B_c \to a_2^+ \pi^0$
& $5.13^{+1.61+0.59+0.94+0.00}_{-1.40-0.56-0.68-0.01}$
  &$B_c \to K_2^{*0} \pi^+$
& $3.73^{+1.66+0.31+0.37+0.04}_{-1.18-0.31-0.44-0.02}$
 \\ \hline
  $B_c \to a_2^0 \pi^+$
& $5.13^{+1.61+0.59+0.94+0.00}_{-1.40-0.56-0.68-0.01}$
  &$B_c \to K_2^{*+} \pi^0$
  &$1.87^{+0.82+0.15+0.18+0.01}_{-0.59-0.16-0.23-0.02}$
 \\ \hline
  $B_c \to a_2^+ \eta$
& $3.92^{+0.39+0.73+0.18+0.35+0.12}_{-0.31-0.70-0.28-0.34-0.11}$
  &$B_c \to K_2^{*+} \eta$
  &$5.37^{+1.77+0.54+0.46+0.04+0.15}_{-0.93-0.55-0.34-0.05-0.15}$
 \\ \hline
  $B_c \to a_2^+ \eta^{\prime}$
& $2.56^{+0.25+0.48+0.11+0.22+0.11}_{-0.20-0.46-0.19-0.23-0.11}$
  &$B_c \to K_2^{*+} \eta^{\prime}$
  &$6.51^{+0.90+1.21+0.42+0.39+0.14}_{-0.34-1.14-0.35-0.37-0.16}$
 \\ \hline
  $B_c \to f_2 \pi^+$
& $7.38^{+0.63+1.59+0.16+0.72}_{-0.39-1.53-0.31-0.70}$
  &$B_c \to a_2^+ K^0$
  &$9.04^{+1.52+1.62+1.11+0.52}_{-0.90-1.45-1.34-0.44}$
 \\ \hline
  $B_c \to K_2^{*+} \bar{K}^0$
& $11.14^{+0.80+1.64+2.42+0.49}_{-0.15-1.45-1.07-0.33}$
  &$B_c \to a_2^0 K^{+}$
  &$4.52^{+0.76+0.82+0.57+0.26}_{-0.45-0.72-0.67-0.21}$
 \\ \hline
  $B_c \to \bar{K}^{*0}_2 K^+$
& $10.46^{+4.82+0.93+0.84+0.20}_{-2.96-0.87-1.50-0.22}$
  &$B_c \to f_2 K^+$
  &$4.56^{+0.72+0.93+0.55+0.30}_{-0.42-0.84-0.65-0.26}$
 \\ \hline
  $ $        & $ $
  &$B_c \to f_2^{\prime} K^+$
  &$6.33^{+2.79+0.40+0.46+0.11}_{-1.82-0.39-0.86-0.15}$
 \\ \hline \hline
\end{tabular}}
\end{center}
\end{table}
Based on the pQCD predictions of the {\it CP}-averaged branching ratios for the considered decay channels $B_c \to TP$ presented in Table~\ref{tab:bc2tp}, one can find the following results:
\begin{itemize}
\item[]{(1)}
Relative to the suppressed CKM matrix element $V_{us}\sim 0.22506$~\cite{Olive:2016xmw} 
in the $\Delta S=1$ modes, the enhanced one $V_{ud} \sim 0.97434$ in the $\Delta S=0$ 
modes makes their decay rates generally much larger 
around one order, which can be clearly seen in Table~\ref{tab:bc2tp}.

\item[]{(2)}
Generally speaking, the nonleptonic charmless $B_c \to TP$ modes have decay rates from $10^{-7}$(e.g., $B_c \to f_2 \pi^+$) to $10^{-8}$(e.g., $B_c \to K_2^{*+} \pi^0$) in the pQCD framework, except for the two $B_c \to K_2^* K$ processes with large branching ratios,
\beq
Br(B_c \to K_2^{*+} \bar{K}^0) &=& 1.11^{+0.31}_{-0.18} \times 10^{-6}\;, \qquad
Br(B_c \to \bar{K}_2^{*0} K^+) = 1.05^{+0.50}_{-0.34} \times 10^{-6}\;,
\eeq
which are expected to be tested in the near future since, as argued in Ref.~\cite{DescotesGenon:2009ja}, the $B_c$ decays with the branching ratios of $10^{-6}$ can be measured at the LHC experiments. In light of the still large theoretical errors in these two modes, we usually provide a more precise ratio between these two {\it CP}-averaged branching ratios $Br(B_c \to  K_2^{*+} \bar{K}^0)$ and $Br(B_c \to \bar{K}_2^{*0} K^+)$ as
\beq
R_{\bar{K}^0/K^+} &\equiv& \frac{Br(B_c \to  K_2^{*+} \bar{K}^0)}{Br(B_c \to \bar{K}_2^{*0} K^+)} \approx
1.07^{+0.40+0.05+0.13+0.02}_{-0.29-0.06-0.00-0.01}\;,
\label{eq:rk2k}
\eeq
in which the uncertainties induced by the hadronic inputs could
be greatly canceled. Of course, the largest error of the ratio $R_{\bar{K}^0/K^+}$ arising from the charm-quark mass in the
$B_c$ meson distribution amplitude $\phi_{B_c}$ indicates that
much more effort should be devoted to better understanding the
nonperturbative QCD dynamics involved in the $B_c$ meson, which
will be helpful to further provide theoretical predictions with
good precision for experiments. Compared to the $B_c \to K_2^* K$
modes, it is worth noticing the different phenomenologies exhibited
in the $B_c \to K^* K$ decays~\cite{Liu:2009qa}. The decay rate
of $B_c \to \bar{K}^{*0} K^+$ is much larger than that of
$B_c \to \bar{K}^0 K^{*+}$ by a factor of about 5.5, in terms of
the central values. The underlying reason is that, relative to
the antisymmetric $K_2^*$ light-cone distribution amplitudes
[see Eq.~(\ref{eq:twist-2})] in the $SU(3)$ limit~\cite{Cheng:2010hn},
the significant $SU(3)$ flavor symmetry-breaking effects have been 
included in both $K$
and $K^*$ mesons, which can be seen evidently from the $a_1$ terms in their 
leading-twist distribution amplitudes~\cite{Liu:2009qa}.

\item[]{(3)}
The $B_c \to TP$ modes involving $\eta-\eta^\prime$ mixing effects [i.e., 
$B_c \to a_2^+ (\eta, \eta')$ and $B_c \to K_2^{*+} (\eta, \eta')$ decays] show 
different interferences between $\eta_q$ and $\eta_s$ flavor states. That is, there 
is constructive(destructive) interference in 
the $B_c \to a_2^+ \eta(B_c \to a_2^+ \eta')$ mode, while the opposite occurs 
in the $B_c \to K_2^{*+} \eta(B_c \to K_2^{*+} \eta')$ channel. Furthermore, one can deduce the dominance of $\eta_q (\eta_s)$ contributions in the $B_c \to a_2^+ \eta^{(\prime)} (B_c \to K_2^{*+} \eta^{(\prime)})$ modes based on the numerical results of the branching ratios displayed in Table~\ref{tab:bc2tp}. Similar interferences have also been observed in the $B_c \to \rho^+ (\eta, \eta')$ and $B_c \to K^{*+} (\eta, \eta')$ decays~\cite{Liu:2009qa}. We explicitly present four interesting ratios among the above-mentioned $B_c \to (\rho, K^*, a_2, K_2^*)(\eta, \eta')$ decays:
\beq
R_{\eta/\eta'}^{a_2}&\equiv& \frac{Br(B_c \to a_2^+ \eta)}{Br(B_c \to a_2^+ \eta')}=1.53^{+0.03}_{-0.02}  \;, \qquad
R_{\eta/\eta'}^{\rho}\equiv \frac{Br(B_c \to \rho^+ \eta)}{Br(B_c \to \rho^+ \eta')}=1.50^{+0.00}_{-0.02}  \;,\\
R_{\eta'/\eta}^{K_2^*}&\equiv& \frac{Br(B_c \to K_2^{*+} \eta')}{Br(B_c \to K_2^{*+} \eta)}=1.21^{+0.22}_{-0.21}  \;, \qquad
R_{\eta'/\eta}^{K^*}\equiv \frac{Br(B_c \to K^{*+} \eta')}{Br(B_c \to K^{*+} \eta)}=4.22^{+0.76}_{-1.59}  \;,
\eeq
where various errors in the ratios have been added in quadrature. The good isospin symmetry makes the $R_{\eta/\eta'}^{a_2}$ approximately equal to the $R_{\eta/\eta'}^{\rho}$; however, the significant $SU(3)$ flavor symmetry-breaking effects in the $K^*$ meson makes the $R_{\eta'/\eta}^{K^*}$ quite different from the $R_{\eta'/\eta}^{K_2^*}$. It is expected that future precise measurements of these ratios might be helpful to investigate the possible pseudoscalar glueball in the $\eta'$ state~\cite{Cheng:2008ss,Liu:2012ib}.

\item[]{(4)}
As far as the $B_c \to (a_2^0, f_2) (\pi^+, K^+)$ channels
are concerned, one can find that, according to the pQCD
predictions for the branching ratios, the
constructive (destructive) interferences between $u\bar u$
and $d\bar d$ components in the $f_2 (a_2^0)$ meson with the 
same (opposite) sign result in a slightly larger (smaller)
$Br(B_c \to f_2 \pi^+)=7.38^{+1.86}_{-1.75}\times 10^{-7} 
[Br(B_c \to a_2^0 \pi^+)=5.13^{+1.96}_{-1.65} \times 10^{-7}]$.
On the other hand, due to only the $u\bar u$ component in both $a_2^0$ and $f_2$ states giving contributions, the almost equivalent branching ratios $Br(B_c \to a_2^0 K^+) \approx Br(B_c \to f_2 K^+)$ can be obtained, which are more like that seen in the $B_c \to (\rho^0, \omega) K^+$ modes~\cite{Liu:2009qa}. The negligibly tiny deviations between the $B_c \to a_2^0 K^+$ and $B_c \to f_2 K^+$ decays arise from the slightly different decay constants and hadron masses of the $a_2^0$ and $f_2$ states, as well as from the same QCD behavior at leading twist. Likewise, the similar phenomenologies of the branching ratios and polarization fractions can be seen clearly from the processes of $B_c \to K_2^{*+} (\rho^0, \omega)$ and $B_c \to (a_2^0, f_2) K^{*+}$ in Table~\ref{tab:bc2tv1}.

\item[]{(5)}
Some simple relations and many other interesting ratios, which can shed light on the (non)validity of $SU(3)$ flavor symmetry in the considered decays, are given as follows:
\beq
Br(B_c \to \bar{K}_2^{*0} \pi^+)&=& 2\cdot Br(B_c \to K_2^{*+} \pi^0)  \non
&=& Br(B_c \to \bar{K}_2^{*0} K^+) \cdot (|\frac{V_{us}}{V_{ud}}| \cdot\frac{f_\pi}{f_K})^2\non
&\sim & Br(B_c \to K_2^{*+} \bar{K}^0)\cdot (|\frac{V_{us}}{V_{ud}}| \cdot \frac{f_\pi}{f_K})^2\;,
\label{eq:su3-k2k-k2pi}
\\
Br(B_c \to a_2^+ K^0)&=& 2\cdot Br(B_c \to a_2^{0} K^+) \;,
\eeq
\beq
R_{K/\pi}^{a_2}&\equiv&\frac{Br(B_c \to a_2^0 K^+)}{Br(B_c \to a_2^0 \pi^+)}=0.088^{+0.024+0.005+0.001+0.005}_{-0.010-0.005-0.004-0.004}\;,
\label{eq:a2-Kpi}\\
R_{K/\pi}^{f_2}&\equiv&\frac{Br(B_c \to f_2 K^+)}{Br(B_c \to f_2 \pi^+)}=0.062^{+0.004+0.001+0.006+0.002}_{-0.003-0.001-0.007-0.002}\;.
\label{eq:f2-Kpi}
\eeq
Moreover, the ratio between $Br(B_c \to f_2 K^+)$ and $Br(B_c \to f'_2 K^+)$ 
when confronted 
with the future precision data can provide useful hints for the $f_2-f'_2$ mixing, though the ideal mixing is assumed in this work,
\beq
R_{f_2/f'_2}^{K}&\equiv&\frac{Br(B_c \to f_2 K^+)}{Br(B_c \to f'_2 K^+)}= 0.72^{+0.20+0.10+0.03+0.03}_{-0.14-0.09-0.01-0.02}\;.
\label{eq:rk-f2}
\eeq
where the largest error of the ratio is also induced by
the variations of the $B_c$ meson distribution amplitude.
Therefore, an in-depth understanding of the hadronization
of the involved meson is the key to provide precise
predictions in the pQCD
approach for future experimental measurements.
\end{itemize}

\begin{table}[hbt]
\caption{ {\it CP}-averaged branching ratios and polarization fractions of CKM-favored $B_c \to TV$ modes in the pQCD approach.}
\label{tab:bc2tv0}
 \begin{center}\vspace{-0.3cm}{
\begin{tabular}[t]{c|c|c|c}
\hline  \hline
  Decay Modes($\Delta S=0$)   &  Branching ratios($10^{-6}$)
&  Polarization fractions $f_L$(\%) &  Polarization fractions $f_T$(\%) \\
\hline \hline
  $B_c \to a_2^+ \rho^0$
& $1.29^{+0.33+0.18+0.04+0.08}_{-0.29-0.16-0.02-0.07}$
  &$78.2^{+4.4+1.0+0.3+1.3}_{-6.3-1.0-0.2-1.2}$
  &$21.8^{+6.3+1.0+0.2+1.2}_{-4.4-1.0-0.3-1.3}$
 \\
 \hline
  $B_c \to a_2^0 \rho^+$
& $1.29^{+0.33+0.18+0.04+0.08}_{-0.29-0.16-0.02-0.07}$
  &$78.2^{+4.4+1.0+0.3+1.3}_{-6.3-1.0-0.2-1.2}$
  &$21.8^{+6.3+1.0+0.2+1.2}_{-4.4-1.0-0.3-1.3}$
 \\
 \hline
  $B_c \to a_2^+ \omega$
& $0.87^{+0.11+0.12+0.03+0.02}_{-0.09-0.13-0.04-0.02}$
  &$97.4^{+0.3+0.3+0.1+0.1}_{-0.3-0.3-0.1-0.1}$
  &$2.6^{+0.3+0.3+0.1+0.1}_{-0.3-0.3-0.1-0.1}$
 \\
 \hline
  $B_c \to f_2 \rho^+$
& $0.97^{+0.11+0.17+0.04+0.00}_{-0.11-0.16-0.04-0.02}$
  &$98.0^{+0.2+0.3+0.1+0.0}_{-0.2-0.4-0.1-0.1}$
  &$2.0^{+0.2+0.4+0.1+0.1}_{-0.2-0.3-0.1-0.0}$
 \\
 \hline
  $B_c \to K_2^{*+} \bar{K}^{*0}$
& $1.75^{+0.02+0.15+0.08+0.10}_{-0.08-0.15-0.07-0.07}$
  &$82.7^{+0.0+0.8+0.4+1.0}_{-0.7-0.7-0.2-0.6}$
  &$17.3^{+0.7+0.7+0.2+0.6}_{-0.0-0.8-0.4-1.0}$
 \\
 \hline
  $B_c \to \bar{K}_2^{*0} K^{*+}$
& $1.54^{+0.57+0.14+0.08+0.12}_{-0.45-0.13-0.07-0.12}$
  &$81.2^{+5.2+0.1+0.1+1.3}_{-7.9-0.4-0.2-1.7}$
  &$18.8^{+7.9+0.4+0.2+1.7}_{-5.2-0.1-0.1-1.3}$
 \\
 \hline \hline
\end{tabular}}
\end{center}
\end{table}

Now we turn to the analyses of the branching ratios and polarization fractions of the $B_c \to TV$ decays in the pQCD approach. As
stressed previously, due to the angular moment conservation, the
$B_c \to TV$ decays contain three helicities, which are more like
the $B_c \to VV$ ones. Then the definitions of the related
helicity amplitudes, polarization fractions, and relative
phases are also the same as those of $B_c \to VV$ modes 
(see Ref.~\cite{Liu:2009qa} for details). It should
be noted that, as this is a first investigation of the nonleptonic
charmless $B_c \to TV$ decays, only {\it CP}-averaged
branching ratios and polarization fractions (whose values are 
collected in Tables~\ref{tab:bc2tv0}
and~\ref{tab:bc2tv1}) are presented in this work.
Moreover, we specify the polarization fractions as
longitudinal $f_L$ and transverse $f_T(=1-f_L)$,
not those $f_L$, $f_\parallel$, and $f_\perp$
adopted previously~\cite{Liu:2009qa}.
Some remarks are in order:
\begin{itemize}
\item[]{(1)}
Differently from the $B_c \to TP$ decays, all of the CKM-favored 
$B_c \to TV$ modes (which contain three polarization contributions with larger decay constants and hadron masses of vector mesons) have decay rates of $10^{-6}$ within theoretical errors in the
pQCD approach at leading order. It is believed that the predictions of these large branching ratios can be confirmed soon by the LHC experiments at CERN~\cite{DescotesGenon:2009ja}. The {\it CP}-averaged branching ratios of the 
CKM-suppressed $B_c \to TV$ modes are nearly $10^{-8}-10^{-7}$, which may have to 
await future tests with much larger data samples. 
Nevertheless, one can easily find that all of the $B_c \to TV$ modes are governed by the longitudinal decay amplitudes, which result in the large polarization fractions in the range of $78\%-98\%$, as presented in Tables~\ref{tab:bc2tv0} and~\ref{tab:bc2tv1}.

\item[]{(2)}
As shown in Table~\ref{tab:bc2tv0}, the {\it CP}-averaged branching ratios and the polarization fractions of $B_c \to a_2^+ \omega$ and $B_c \to f_2 \rho^+$ channels are close to each other. The reason is that, on the one hand, the pure $\frac{u\bar u+ d\bar d}{\sqrt{2}}$ component for the $f_2(1270)$ state is assumed which is same as the $\omega$ meson with ideal mixing, and on the other hand, the adopted decay constants and masses of the involved tensor and vector mesons are similar in magnitude with only small differences. More specifically,
\beq
m_\rho&=& 0.770~{\rm GeV}\;, \qquad m_\omega= 0.782~{\rm GeV}\;,\qquad f_\rho= 0.209 \pm 0.002~{\rm GeV}\;, \qquad
f_\omega= 0.195 \pm 0.003~{\rm GeV}\;; \non
f_\rho^T&=& 0.165 \pm 0.009~{\rm GeV}\;, \qquad f_\omega^T= 0.145 \pm 0.010~{\rm GeV}\;
\eeq
for the light vector $\rho$ and $\omega$ mesons, and
\beq
m_{a_2}&=& 1.318~{\rm GeV}\;, \qquad m_{f_2}= 1.275~{\rm GeV}\;,\qquad f_{a_2}= 0.107 \pm 0.006~{\rm GeV}\;, \qquad
f_{f_2}= 0.102 \pm 0.006~{\rm GeV}\;; \non
f_{a_2}^T&=& 0.105 \pm 0.021~{\rm GeV}\;, \qquad f_{f_2}^T= 0.117 \pm 0.025~{\rm GeV}\;
\eeq
for the light tensor $a_2$ and $f_2$ states. Therefore, it is also understandable that these two decay rates are a bit smaller than that of the $B_c \to \rho^+ \omega$ channel~\cite{Liu:2009qa}.

\item[]{(3)}
It is interesting to note that the $B_c \to a_2^+ \rho^0$ and $B_c \to a_2^+ \omega$ decay rates indicate different interferences between the $u\bar u$ and $d\bar d$ components in the  $\rho^0$ and $\omega$ mesons. As can be seen in Table~\ref{tab:bc2tv0}, it is evident that the constructive(destructive) interferences contribute to the former(latter) mode. Similar phenomenologies also appear in the $B_c \to a_2^0 \rho^+$ and $B_c \to f_2 \rho^+$ decays. Moreover, one can easily observe that the numerical results of the branching ratios are sensitive to the hadronic parameters such as the charm-quark mass, the decay constants of the light tensor meson, etc. We thus define some ratios among the branching ratios as follows:
\beq
R_{\omega/\rho^0}^{a_2}&\equiv& \frac{Br(B_c \to a_2^+ \omega)}{Br(B_c \to a_2^+ \rho^0)} = 0.67^{+0.11+0.00+0.01+0.03}_{-0.07-0.02-0.02-0.02}\;,\\
R_{f_2/a_2^0}^{\rho}&\equiv& \frac{Br(B_c \to f_2 \rho^+)}{Br(B_c \to a_2^0 \rho^+)} = 0.75^{+0.11+0.03+0.01+0.03}_{-0.08-0.03-0.02-0.04}\;,\\
R_{a_2\omega/f_2\rho}&\equiv& \frac{Br(B_c \to a_2^+ \omega)}{Br(B_c \to f_2 \rho^+)} = 0.90^{+0.01+0.01+0.00+0.02}_{-0.00-0.03-0.01-0.01}\;,
\eeq
where the uncertainties arising from the errors of the inputs
have been greatly canceled, though these parameters involved in
the meson wave functions are not factored out. These ratios and the detectable decay rates could be helpful to further explore the QCD behavior of the $a_2$ and $f_2$ states.

\item[]{(4)}
Analogous to the $B_c \to K_2^* K$ decays, the $B_c \to K_2^{*+}
\bar{K}^{*0}$ and $B_c \to \bar{K}_2^{*0} K^{*+}$ modes also have
 branching ratios that are close to each other for the same reason.
More interestingly, the ratio arising from $Br(B_c \to K_2^{*+}
\bar{K}^{*0})$ over $Br(B_c \to \bar{K}_2^{*0} K^{*+})$ in
the pQCD approach is approximately equal to that [see Eq.~(\ref{eq:rk2k})] 
obtained in the $B_c \to K_2^* K$ decays,
although these two branching ratios
induced by three polarizations are clearly larger than
the $B_c \to K_2^* K$ ones
only from longitudinal polarization. The related branching
ratios and ratio are,
\beq
Br(B_c \to  K_2^{*+} \bar{K}^{*0})&=&1.75^{+0.20}_{-0.20}\times 10^{-6} \;, \qquad
Br(B_c \to  K_2^{*+} \bar{K}^{*0})= 1.54^{+0.60}_{-0.49}\times 10^{-6} \;,
\eeq
and
\beq
R_{\bar{K}^{*0}/K^{*+}}&=&\frac{Br(B_c \to  K_2^{*+} \bar{K}^{*0})}{Br(B_c \to  K_2^{*+} \bar{K}^{*0})}=1.14^{+0.39+0.00+0.00+0.04}_{-0.32-0.01-0.01-0.03}\;.
\eeq
The conservation law of angular momentum results in the tensor $K_2^*$ state contributing to the $B_c \to K_2^* K^*$ decays with only three helicities, $\lambda=0$ and $\pm 1$, which makes it behave more like a vector meson. Since the smaller decay constants (as shown in Table~\ref{tab:m&f-tensor}) of both longitudinal and transverse polarizations of the $K_2^*$ meson are adopted, the decay rates and polarization fractions of the $B_c \to K_2^{*} K^{*}$ modes are basically consistent with those of the $B_c \to \bar{K}^{*0} K^{*+}$ one within errors~\cite{Liu:2009qa}, though $m_{K_2^*}$ is nearly 2 times larger than $m_{K^*}$.

\end{itemize}
\begin{table}[hbt]
\caption{ Same as Table~\ref{tab:bc2tv0} but for CKM-suppressed
$B_c \to TV$ modes.}
\label{tab:bc2tv1}
 \begin{center}\vspace{-0.3cm}{
\begin{tabular}[t]{c|c|c|c}
\hline  \hline
  Decay Modes($\Delta S=1$)   &  Branching ratios($10^{-8}$)
&  Polarization fractions $f_L$(\%) &  Polarization fractions $f_T$(\%) \\
\hline \hline
  $B_c \to K_2^{*0} \rho^+$
& $7.17^{+2.67+0.65+0.14+0.45}_{-2.11-0.62-0.13-0.47}$
  &$84.8^{+4.3+0.2+0.1+0.9}_{-6.5-0.2-0.0-1.1}$
  &$15.2^{+6.5+0.2+0.0+1.1}_{-4.3-0.2-0.1-0.9}$
 \\
 \hline
  $B_c \to K_2^{*+} \rho^0$
& $3.59^{+1.33+0.31+0.07+0.22}_{-1.06-0.31-0.07-0.24}$
  &$84.8^{+4.3+0.2+0.1+0.9}_{-6.5-0.2-0.0-1.1}$
  &$15.2^{+6.5+0.2+0.0+1.1}_{-4.3-0.2-0.1-0.9}$
 \\
 \hline
  $B_c \to K_2^{*+} \omega$
& $3.14^{+1.17+0.28+0.10+0.20}_{-0.92-0.27-0.09-0.20}$
  &$84.4^{+4.4+0.2+0.1+1.0}_{-6.5-0.2-0.0-1.0}$
  &$15.6^{+6.5+0.2+0.0+1.0}_{-4.4-0.2-0.1-1.0}$
 \\
 \hline
  $B_c \to K_2^{*+} \phi$
& $11.80^{+0.15+1.11+0.41+0.26}_{-0.83-1.05-0.38-0.29}$
  &$80.2^{+0.0+1.1+0.3+0.4}_{-1.4-1.2-0.3-0.5}$
  &$19.8^{+1.4+1.2+0.3+0.5}_{-0.0-1.1-0.3-0.4}$
 \\
 \hline
  $B_c \to a_2^+ K^{*0}$
  & $9.20^{+0.00+1.11+0.38+0.26}_{-0.68-1.06-0.38-0.38}$
  &$87.4^{+0.0+0.9+0.2+0.4}_{-0.9-1.0-0.3-0.6}$
  &$12.6^{+0.9+1.0+0.3+0.6}_{-0.0-0.9-0.2-0.4}$
 \\
 \hline
  $B_c \to a_2^{0} K^{*+}$
& $4.60^{+0.00+0.55+0.19+0.12}_{-0.34-0.53-0.19-0.19}$
  &$87.4^{+0.0+0.9+0.2+0.4}_{-0.9-1.0-0.3-0.6}$
  &$12.6^{+0.9+1.0+0.3+0.6}_{-0.0-0.9-0.2-0.4}$
 \\
 \hline
  $B_c \to f_2 K^{*+}$
& $4.38^{+0.00+0.58+0.17+0.09}_{-0.30-0.56-0.20-0.19}$
  &$88.6^{+0.0+1.0+0.2+0.2}_{-0.6-1.1-0.3-0.5}$
  &$11.4^{+0.6+1.1+0.3+0.5}_{-0.0-1.0-0.2-0.2}$
 \\
 \hline
  $B_c \to f_2^{\prime} K^{*+}$
& $9.60^{+3.42+0.66+0.46+0.77}_{-2.68-0.58-0.43-0.69}$
  &$79.4^{+5.6+0.3+0.1+1.5}_{-8.1-0.1-0.0-1.6}$
  &$20.6^{+8.1+0.1+0.0+1.6}_{-5.6-0.3-0.1-1.5}$
 \\
 \hline \hline
\end{tabular}}
\end{center}
\end{table}
\begin{itemize}
\item[]{(5)}
As reported by the {\it BABAR} Collaboration, the fact that 
$f_L/f_T \gg 1$ for $B \to \phi K_2^*$ decays~\cite{Aubert:2008bc} 
while $f_L/f_T \sim 1$ for $B \to \omega K_2^*$ decays~\cite{Aubert:2009sx} 
make the well-known ``polarization puzzle" more confusing, although both of 
them have the same helicity structure as the $B \to \phi K^*$ modes with 
the penguin-dominated contributions. Furthermore, the branching ratio of 
the $B^+ \to \omega K_2^{*+}$ channel is much larger than that of the 
$B^+ \to \phi K_2^{*+}$ one by a factor of around 2.5, which is contrary 
to the ratio of the $B^+ \to \omega K^{*+}$ and $B^+ \to \phi K^{*+}$ 
decay rates~\cite{Olive:2016xmw,Amhis:2016xyh}. The current theoretical 
studies on these anomalous phenomena cannot give satisfactory explanations, 
which means that more investigations on the light tensor $K_2^*$ meson are demanded. 
The {\it CP}-averaged branching ratios and polarization fractions of $B_c \to K_2^* (\rho, \omega, \phi)$ channels are given in the pQCD approach at leading order are presented in Table~\ref{tab:bc2tv1}. One can find that these four modes are dominated by the longitudinal decay amplitudes, and the $B_c \to K_2^{*0} \rho^+$ and $B_c \to K_2^{*+} \phi$ decay rates are on the order of $10^{-7}$ within errors; these are expected to be tested by the LHC Run-II experiments at CERN in the near future.

\item[]{(6)}
Likewise, some interesting ratios of $B_c \to TV$ decays can also provide useful hints about the QCD dynamics involved in the light tensor mesons, as well as in the related decay channels. For example, future precise measurements can tell us the mixing information about $f_2(1275)$ and $f'_2(1525)$ states through the ratio $R^{K^*}_{f_2/f'_2}$,
\beq
R^{K^*}_{f_2/f'_2}&\equiv&\frac{Br(B_c \to f_2 K^{*+})}{Br(B_c \to f'_2 K^{*+})}=0.46^{+0.13+0.02+0.00+0.01}_{-0.12-0.04-0.01-0.03}\;,
\label{eq:rks-f2}
\eeq
where the $B_c \to f'_2 K^{*+}$ branching ratio reaches $10^{-7}$.
If the future measured ratios $R^K_{f_2/f'_2}$ and $R^{K^*}_{f_2/f'_2}$
 deviate from those predicted in  
Eqs.~(\ref{eq:rk-f2}) and~(\ref{eq:rks-f2}), then
the mixture of the $f_{2q}$ and $f_{2s}$ flavor states
should be included for the $f_2$ and $f'_2$ mesons.
It is noted that the $R^{K}_{f_2/f'_2}$ is a bit larger than the
$R^{K^*}_{f_2/f'_2}$ by a factor of around 1.5, since
$Br(B_c \to f_2 K^+) \sim Br(B_c \to f_2 K^{*+})$ while
$Br(B_c \to f'_2 K^{*+}) \sim 1.5 \times Br(B_c \to f'_2 K^+)$.
More data are demanded on the $f_2$ and $f'_2$ states to
further understand these phenomenologies, in particular,
the approximately equal
decay rates between $B_c \to f_2 K^+$ and $B_c \to f_2 K^{*+}$ modes,
\beq
Br(B_c \to f_2 K^+)&=&4.56^{+1.33}_{-1.17}\times 10^{-8}  \;, \qquad
Br(B_c \to f_2 K^{*+})=4.38^{+0.61}_{-0.69}\times 10^{-8}\;.
\eeq
After all, the latter process receives contributions from three helicities.

\item[]{(7)}
The isospin symmetry can be observed in the pQCD calculations for $B_c \to K_2^* \rho$ and $B_c \to a_2 K^*$ modes, that is,
\beq
Br(B_c \to K_2^{*0} \rho^+) &=& 2 \cdot Br(B_c \to K_2^{*+} \rho^0)\;,
\qquad
Br(B_c \to a_2^+ K^{*0}) = 2 \cdot Br(B_c \to a_2^0 K^{*+})\;.
\eeq
However, the $SU(3)$ flavor symmetry cannot be easily seen in the $B_c \to K_2^* K^*$ and $B_c \to K_2^* \rho$ decays [Eq.~(\ref{eq:su3-k2k-k2pi})], since these two decays have three helicity structures with different decay constants and different Gegenbauer moments of the vector $K^*$ and $\rho$ mesons in longitudinal and transverse polarizations, respectively. Nevertheless, we can still present the ratios between $Br(B_c \to K_2^* \rho^+)$ and $Br(B_c \to K_2^* K^*)$, which can be used to 
show 
the (non)validity of the $SU(3)$ flavor symmetry by combining future precise measurements,
\beq
R_{\rho/\bar{K}^{*}}&\equiv&
\frac{Br(B_c \to K_2^{*0} \rho^+)}{Br(B_c \to K_2^{*+}\bar{K}^{*0})} = 0.041^{+0.015+0.000+0.001+0.000}_{-0.011-0.000-0.001-0.001}\;, \\
R_{\rho/K^{*}}&\equiv&
\frac{Br(B_c \to K_2^{*0} \rho^+)}{Br(B_c \to \bar{K}_2^{*0}K^{*+})} = 0.047^{+0.000+0.000+0.001+0.000}_{-0.001-0.001-0.002-0.001}\;.
\eeq
Two more relations can also be written as follows:
\beq
R^{a_2}_{K^*/\rho}&\equiv& \frac{Br(B_c \to a_2^0 K^{*+})}{Br(B_c \to a_2^0 \rho^+)} = 0.036^{+0.007+0.000+0.000+0.000}_{-0.008-0.001-0.001-0.002}\;,
\label{eq:a2-ksrho}\\
R^{f_2}_{K^*/\rho}&\equiv& \frac{Br(B_c \to f_2 K^{*+})}{Br(B_c \to f_2 \rho^+)} = 0.045^{+0.002+0.002+0.000+0.001}_{-0.004-0.001-0.000-0.001}\;,
\label{eq:f2-ksrho}
\eeq
where, by combining the ratios $R^{a_2}_{K/\pi}$ and $R^{f_2}_{K/\pi}$ in Eqs.~(\ref{eq:a2-Kpi}) and~(\ref{eq:f2-Kpi}), 
$Br(B_c \to a_2^0 \rho^+) \sim 2.5 \times Br(B_c \to a_2^0 \pi^+)$ but $Br(B_c \to a_2^0 K^{*+}) \simeq Br(B_c \to a_2^0 K^+)$ result in the relation $R^{a_2}_{K/\pi} > R^{a_2}_{K^*/\rho}$. However, $Br(B_c \to f_2 \rho^+) \sim 1.3 \times Br(B_c \to f_2 \pi^+)$ but $Br(B_c \to f_2 K^{*+}) \simeq Br(B_c \to f_2 K^+)$ leads to the relation $R^{f_2}_{K^*/\rho} < R^{f_2}_{K/\pi}$.
It is worth mentioning that the $R^{a_2}_{K/\pi}$ is a bit larger than the $R^{f_2}_{K/\pi}$ in the $B_c \to (a_2^0, f_2)(\pi^+, K^+)$ decays, while the $R^{a_2}_{K^*/\rho}$ is slightly smaller than the $R^{f_2}_{K^*/\rho}$ in the $B_c \to (a_2^0, f_2)(\rho^+, K^{*+})$ modes, which could be tested and further clarified by the related experiments with good precision in the future.

\item[]{(8)}
Recently, the three-body( or quasi-two-body) $B$ meson decays have attracted more and more attention, since the two $B$ factories and the LHC experiments have collected lots of data on the related channels. It is suggested that the considered light tensor states in this work can also be studied through the resonant contributions in the relevant three-body modes~\cite{Aaij:2016qnm}; for example, the $f_2(1270)$ and $f'_2(1525)$ mesons can be investigated in the $B_c \to f_2(1270)(\to \pi\pi) (\pi, K)$ and $B_c \to f'_2(1525)(\to KK) (\pi, K)$ channels, respectively, which can play important roles in exploring the QCD dynamics of the light tensor mesons. These studies can also help to further deepen our understanding of the three-body decay mechanism.

\end{itemize}

In summary, we have analyzed the nonleptonic charmless $B_c \to TP, TV$ decays in the pQCD approach. Due to the angular momentum conservation, the light tensor meson can only contribute with one($\lambda=0$) or three helicities($\lambda= 0, \pm 1$). By properly redefining the polarization tensor, the new polarization vector $\epsilon_T$ of the light tensor meson can be obtained, which is slightly different than the $\epsilon_V$ of the vector meson with coefficients $\sqrt{\frac{2}{3}}$ and $\sqrt{\frac{1}{2}}$ for longitudinal and transverse polarizations, respectively. Therefore, the decay amplitudes can be easily presented with appropriate replacements from the $B_c \to PV, VV$ decay modes.
The {\it CP}-averaged branching ratios and polarization fractions for the considered channels have been predicted in the pQCD approach. Most of the CKM-favored processes have decay rates of $10^{-6}$, which are expected to be measured soon by the LHC experiments at CERN. Numerically, all of the $B_c \to TV$ modes are governed by the longitudinal contributions. Many interesting ratios among the branching ratios have been derived, as well as some simple relations that can be used to exhibit the (non)validity of the $SU(3)$ flavor symmetry. The predictions about these concerned $B_c \to TP, TV$ decays in the pQCD approach can be confronted with measurements in the (near) future, which are expected to shed some light on the annihilation decay mechanism in the related decay channels.

\bigskip

This work is supported by the National Natural Science
Foundation of China under Grants No.~11205072, No.~11235005,
No.~11447032, No.~11505098, and No.~11647310 
and by the Research Fund of Jiangsu Normal University under Grant No.~HB2016004, by the Doctoral Scientific Research Foundation of Inner Mongolia University,
and by the Natural Science Foundation of Shandong Province
under Grant No.~ZR2014AQ013.


\end{document}